\documentclass{ws-p8-50x6-01}
\usepackage{amssymb}
\usepackage{amsmath}
\begin{document}
\thispagestyle{empty}

\title{RAPIDITY GAPS IN QUARK AND GLUON JETS -\\
A PERTURBATIVE APPROACH\footnote{to appear in 
Proc. XXIX International Symposium
QCD \& Multiparticle Production (ISMD99), Brown University, Providence, RI,
USA, Aug. 1999, MPI-Pht 99-49}}

\author{Wolfgang Ochs}

\address{Max-Planck-Institut f\"ur Physik, 
(Werner-Heisenberg-Institut)\\ 
F\"ohringer Ring 6, D-80805 M\"unchen, Germany
}
\author{Tokuzo Shimada}

\address{Meiji University, Departement of Physics, 
Higashi Mita 1-1, Tama\\
Kawasaki, Kanagava 214, Japan
}


\maketitle

\abstracts{
We derive the probability for  rapidity gaps 
in a parton cascade 
and investigate the dual connection with hadronic final states.
A good description
of observations in $e^+e^-$-annihilations is obtained by perturbative
QCD calculations in MLLA
using previously determined parameters (QCD scale $\Lambda$ and  $k_T$-cut-off
$Q_0$) and applying the parton hadron duality picture.
Further predictions are derived; especially, for gaps between jets at 
variable resolution we predict 
 a strong variation of gap probabilities for small
parameters $y_{cut}\to 0$ in the transition from jets to
hadrons.
Large gaps between partons correspond to  large spatial separations of colour
charges: a colour blanching mechanism by soft processes is suggested.
}

\section{Introduction}
The occurrence of large rapidity gaps in  $e^+e^-$-annihilations, or more
generally, inside quark and gluon jets, provides an 
interesting testing ground for
models of colour confinement and hadronization. 

In case of
$e^+e^-$-annihilation the primary process is the production of a
$q\overline q$ pair. According to a perturbative mechanism \cite{bbh,er}
large rapidity gaps in the hadronic final state occur if
in a subsequent process two 
low mass parton pairs are formed in  colour singlet states  
($q\overline q$ or $gg$) 
recoiling against each other.
 The production rates predicted for these ``hard''
colour singlets which involve highly virtual intermediate gluons, however,
are much smaller than those observed by SLD,\cite{SLD} by about two orders of
magnitude.
 
A quantitative description of the data is provided by the JETSET MC 
\cite{jetset} which combines an initial
parton cascade, cut off at a scale $\sim$ 1 GeV, with a 
string hadronization model.
The SLD data are shown in Fig. 1a. The JETSET result fits the data in
the full range of rapidity intervals $\Delta y$ \cite{SLD}; for large 
 $\Delta y$ the contribution of
$\tau^+\tau^-$ events becomes important and the dashed histogram represents the
purely hadronic contribution.
\begin{figure}[t]
\begin{center}
\noindent
\begin{minipage}{5cm}
\mbox{\epsfig{bbllx=2.6cm,bblly=8.2cm,bburx=13.4cm,bbury=23.2cm,%
file=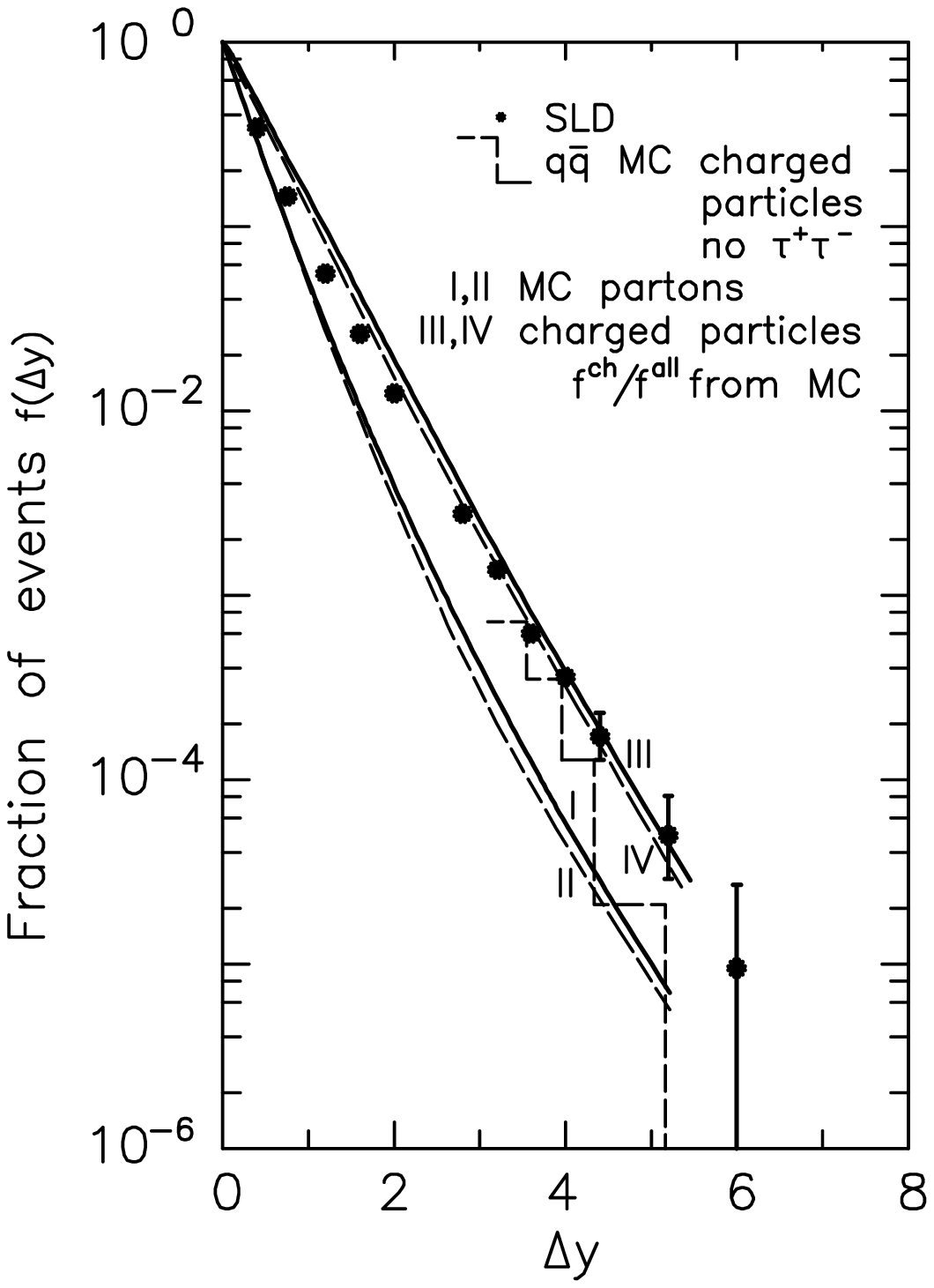,
width=5.0cm}}
%
\end{minipage}
\hspace{1.0cm}
\begin{minipage}{5cm}
\mbox{\epsfig{bbllx=3.8cm,bblly=8.3cm,bburx=14.5cm,bbury=23.2cm,%
file=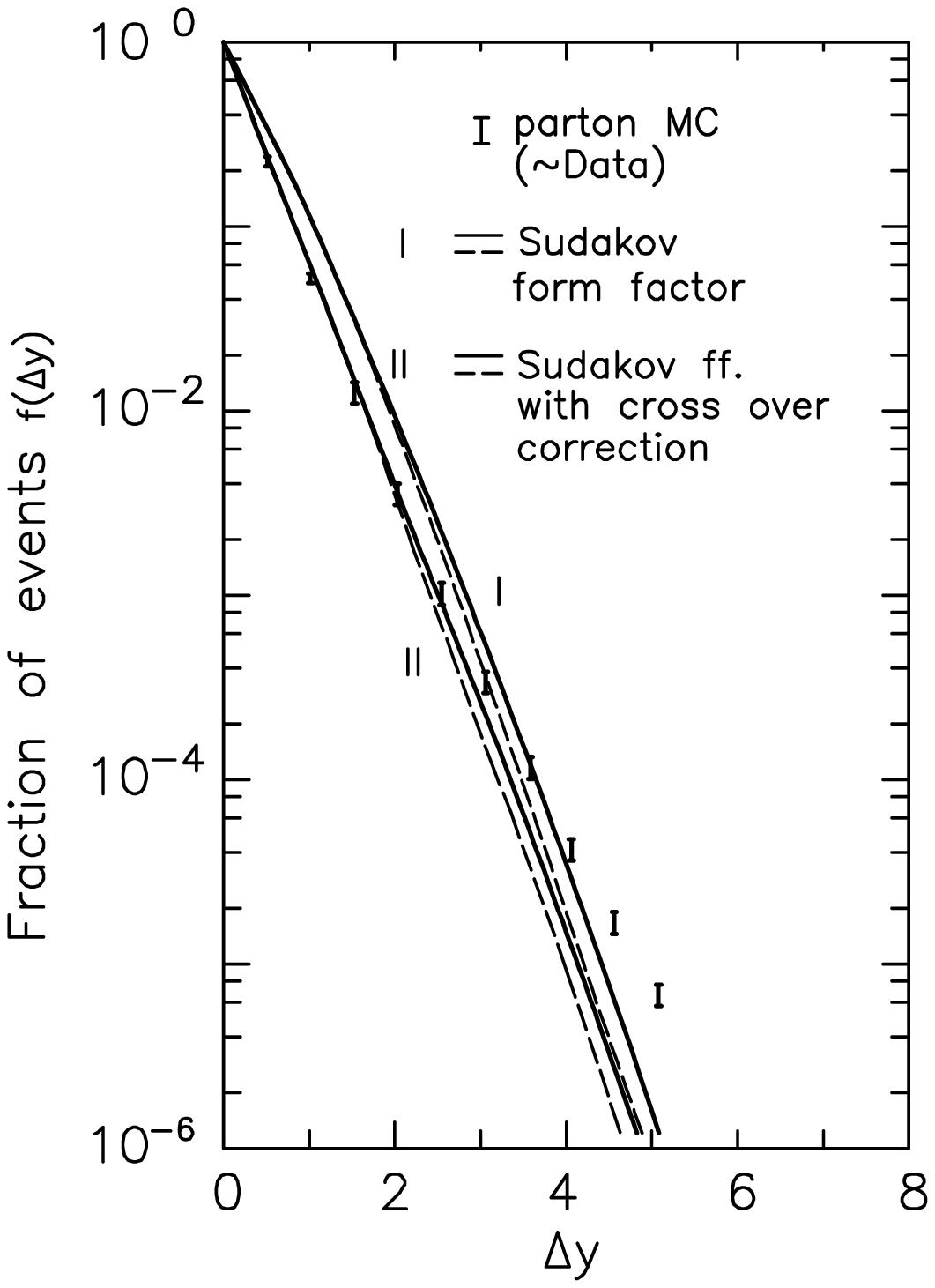,
width=5.0cm}}
%
\end{minipage}
\end{center}
\caption{
Probability for rapidity gaps of length $\Delta y$; (a)
between charged particles in $e^+e^-$ annihilation: data from
SLD\protect\cite{SLD}, also shown is
 the JETSET MC for
$q\overline q$ initial state without $\tau^+\tau^-$ events.\protect\cite{SLD} 
Furthermore we show our results from ARIADNE MC
at the parton level which, in the duality picture, correspond to gaps between
all hadrons (curves I, II); an estimate for gaps between charged particles
is obtained by multiplying these results with the ratio of gap fractions
$f^{ch}/f^{all}$ from the full MC after string hadronization (curves
III,IV) (Parameters used in MC: 
(I) $\Lambda=0.20$ GeV, $\lambda=0.015$, N$_f=3$;    
(II) $\Lambda=0.32$ GeV, $\lambda=0.015$, N$_f=5$).
The data points\protect\cite{SLD} are moved to the right edge of
every interval as appropriate for a cumulative quantity;
(b) vertical bars represent the same ARIADNE results (curves I,II in (a))
which are also close to the data; 
the curves (I) here correspond to our analytical calculation of the Sudakov
form factor, the curves (II) include also the cross over correction
(full curves: $\Lambda=500$ MeV, Durham $k_T$;
dashed:  $\Lambda=350$ MeV, standard $k_T$; $\lambda=0.015$ always)
 \label{fig:data}}
\end{figure}

\enlargethispage{1.2cm}

We will discuss here another approach, originally proposed for the treatment
of single inclusive spectra (``Local Parton Hadron Duality''-
LPHD\cite{lphd}) 
\pagebreak
\noindent
in which the parton cascade is 
perturbatively evolved further 
with cutoff  $k_T>Q_0$ at $Q_0\gtrsim\Lambda$
with QCD scale $\Lambda\sim$ few 100 MeV.
The results are compared directly to the data without any
explicit hadronization phase and the partonic final state is assumed 
to represent
 the hadronic one in a dual sense. This simple approach 
has been applied to a variety of problems,
mainly on inclusive quantities, with rather surprising successes.\cite{dkmt,ko} 

Our present  study carries this idea further, in that we consider an observable
which tends to become exclusive in the limit of large gaps. Indeed,
if colour charges are separated by a large gap, then
one might expect, according to conventional wisdom, a neutralization 
 by non-perturbative processes leading to large deviations from the
perturbative calculations. At the time of a previous study 
along these lines\cite{os} the ratio  $Q_0/\Lambda$ was not well determined
and only an upper limit of the
gap fraction has been given. Meanwhile, from an 
improved analysis of jet and hadron
multiplicities\cite{lo} a determination of this ratio has
been obtained. 
Now, the gap rate can be predicted in absolute terms within the
perturbatively based duality picture.    

\section{Rapidity Gaps in the Perturbative Parton Cascade}
The probability for no radiation into a certain angular interval
is given in field theory by the
exponential Sudakov form factor \cite{suda}, originally derived in QED.
In our application we consider the rapidity gap without gluons 
above the transverse momentum cutoff $Q_0$.
Let us consider specifically the angular interval between $\Theta_1$ and $
\Theta_2$ ($\Theta_1 > \Theta_2$), 
the rapidity is then obtained
from $y = -\ln {\rm tg} \frac{\Theta}{2}$. 
Let us further denote the probability for emission of a gluon
at an angle $\Theta^\prime$ with the energy $\omega^\prime$
off a parent parton $p$ (either a gluon(g) or a quark(q))
as                        
$ \wp_p(\omega^\prime,\Theta^\prime)=
dn_p/d\omega^\prime d\Theta^\prime$.
Then, the Sudakov form factor for the 
angular ordered cascade is given by
\begin{eqnarray}          
\Delta_p(P, \Theta,Q_0) &=& \exp (-w_p(P,\Theta,Q_0)) \label{Delta} \\
w_p(P,\Theta,Q_0) &=& \int d\omega^\prime \int_{k_\perp>Q_0} d\Theta^\prime
\wp_p(\omega^\prime,\Theta^\prime),
\label{wp}
\end{eqnarray}            
and it represents the probability for no gluon being emitted
within the cone of half angle $\Theta$ at
transverse momentum above $Q_0$
from the parent parton $p$
of energy $P=Q/2$.
In particular,
\begin{eqnarray}
\Delta_p(\Theta_2)/\Delta_p(\Theta_1)
= \exp \left( -w_p(\Theta_2)+w_p(\Theta_1) \right)
\end{eqnarray}
represents the probabilty that there is no emission of a gluon
with emission angle between $\Theta_1$ and $\Theta_2$.
These rates have been calculated in
different approximations\\

\noindent
{\em The double logarithmic approximation (DLA)}\\
The simplest approximation takes into account only the leading
contributions from the angle and energy singularities of the gluon emission.
The gap probability $f_p$ is easily calculated analytically.\cite{os} 
For the symmetrical gap in the $cms$ we find a good approximation
for not too large gaps $\Delta y/2 \ll Y$
\begin{gather}
f_p(\Delta y)\simeq \exp(-A_p\Delta y)  \label{dlaslope} \\
    A_p=\frac{4C_p}{b}\ln\frac{Y}{\lambda},
\hspace{0.5cm}    Y=\ln\frac{P\Theta}{Q_0},
\hspace{0.5cm}   \lambda=\ln\frac{Q_0}{\Lambda},
\hspace{0.5cm}     b=\frac{11}{3}N_C-\frac{2}{3} N_f 
   \label{defs}
\end{gather}
with $C_g=3,\ C_q=\frac{4}{3}$.
The gap rate decreases exponentially  with $\Delta y$.
The slope depends sensitively on $\lambda$ for small $\lambda$
and $A\to \infty$ for $\lambda\to 0$.

\noindent
{\em The modified leading logarithmic approximation (MLLA)}\\
In this improved approximation also the next to leading logarithmic terms
are included. Some analytic results have been obtained before.\cite{os}
Here we calculate the probability $w_q$ as in the multiplicity analysis
\cite{lo}   
\begin{equation}
w_q=\int_{Q_0}^{\kappa} d\kappa' \int_{Q_0/\kappa'}^{1-Q_0/\kappa'} dz
    \frac{\alpha_s(k_T)}{2\pi} \Phi_{qg}(z)    \label{wpmlla}
\end{equation}
by numerical integration,
where $\kappa=Q\sin(\Theta/2)$ is the jet virtuality at opening angle
$\Theta$, the splitting function is $\Phi_{qg}(z) = 2 C_F (1+(1-z)^2)/z$
and $k_T$ denotes the transverse momentum; this is taken
as $k_T=z(1-z)\kappa$ (``standard'') or $k_T=\min(z,1-z)\kappa$ (``Durham'').
The different $k_T$ lead to somewhat different $\Lambda$ without changing
$\lambda=0.015$.

In this calculation the exponent $w_q$ is of ${\cal O} (\alpha_s)$. To next
order, processes play a role where a secondary gluon is emitted into the gap
although the firstly emitted gluon is outside the gap, this we call
``cross-over effect''. We note that there is no such effect for
$\Theta'<\Theta_2/2$ because of angular ordering,
therefore, the maximal effect is a shift by $\Delta y=\ln 2\sim 0.7$.
Otherwise, we take the
effect into account by a correction factor in DLA accuracy: we multiply
(\ref{wp}) under the integral
with the probability $P$ for no secondary emission into the gap,
approximately with
\begin{align}
 P = & \exp(-\frac{1}{2}(w_g(k',\Theta')-w_g(k',\Theta_2-\Theta')))
  \quad \textrm{for}\quad \frac{\Theta_2}{2} < \Theta' < \frac{\Theta_1}{2} 
  \label{cr1}\\
 P = & \exp(-\frac{1}{2}(w_g(k',\Theta_1-\Theta')-w_g(k',\Theta_2-\Theta')))
 \quad \textrm{for}\quad  \Theta' > \frac{\Theta_1}{2}
  \label{cr2}
\end{align}

\noindent
{\em The Parton Monte Carlo}\\
As a control of our analytical calculations we compare also with 
the  ARIADNE MC\cite{ariadnep} at parton level which is
based on similar principles, i.e. cutoff $k_T>Q_0$ and possibility to choose
a small $\lambda$ parameter.
We take
parameters (I) determined from a fit to hadron multiplicities\cite{low}; 
new parameters (II) are determined to improve the fit to jet multiplicities at
small $y_{cut}$. In the MC we used only $u$-quarks with $m_u=0$ and kept
number of flavours $N_f$ fixed.

\section{Comparison with Data}
The MC results for the two sets of parameters (I) and (II) 
are also shown in Fig. 1a. They refer to the gaps between all hadrons in the 
duality picture. To obtain an estimate for the gaps between charged hadrons
only we multiply these curves with the respective ratio
derived from the parton and full hadron MC 
which we parametrized as 
$f^{ch}/f^{all}=1+2\Delta y - 0.1 (\Delta y)^2$. 
After this correction,
one observes a very good agreement of this 2-parameter model
with the data (at large $\Delta y$
one should compare to the dashed line). 

In Fig. 1b these MC results  are represented again as vertical
bars for reference, also to the data. The curves (I)  represent our
analytical calculations based on (\ref{wpmlla}) for two sets of 
parameters\cite{lo},
curves (II) include also the cross over corrections (\ref{cr1}),(\ref{cr2}).
Good agreement of the latter results with the MC 
is obtained up to $\Delta y\sim 3-4$, it falls below the MC at higher $\Delta y$
but could still be close to the experimental data.  
\begin{figure}[t]
\begin{center}                                                              
\epsfxsize=5cm
\epsfbox{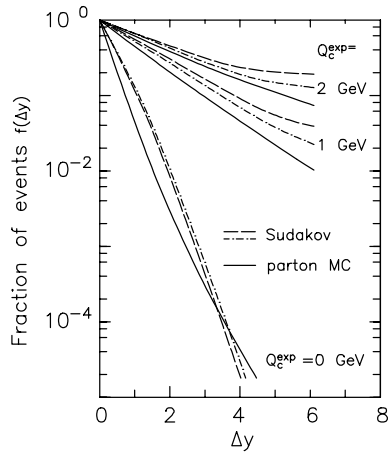} 
\end{center}
\caption{
Probability for rapidity gaps of size $\Delta y$ between jets for 
different resolution parameters $y_{cut}=Q_c^{exp}/s$ in the Durham algorithm,
so $Q_c^{exp}=0$ corresponds to full resolution, i.e. the hadronic final
state. The full line represents ARIADNE MC (curve II in Fig. 1),
the others the analytical Sudakov calculations without crossover effect
(dashed:  $\Lambda=350$ MeV, standard $k_T$;
 dash-point: $\Lambda=500$ MeV, Durham $k_T$,
 $\lambda=0.015$ always, Q=91 GeV)
 \label{fig:ycut}}
\end{figure}

\section{Further Predictions}
We note three consequences of our approach which follow directly from the
simple DLA formula (\ref{dlaslope}).\\
\noindent {\em Quark vs. gluon jet}\\
The slope is proportional to $C_p$ as the Sudakov form factor is derived from
the ${\cal O} (\alpha_s)$ gluon emission probability. 
Then, in DLA, the slope in a gluon jet
is larger by $C_A/C_F=9/4$ as compared to a quark jet.\\
\noindent {\em Energy dependence}\\
The slope behaves like $A\sim \ln\ln (P/Q_0)$ so the gap distribution gets
steeper with increasing jet energy $P$.\\
\noindent {\em Dependence on cutoff $Q_0$ and jet resolution}\\
In the duality picture, 
the cutoff $Q_0$ appears 
as a hadronization scale
which limits the resolution of separate partons. In the evolution equation
it can be interpreted also as jet resolution parameter in the Durham
algorithm whith $y_{cut}=(Q_{cut}/Q)^2$ with $Q_0$ replaced by $Q_{cut}$. Therefore, we
expect a strong dependence of the slope on the jet resolution parameter
through $A\sim -\ln\lambda$.
In the theoretical calculation all hadrons are resolved for $Q_{cut}\to
Q_0$ in the duality picture, experimentally for $Q_{cut}\to 0$. This
mismatch can be resolved by relating\cite{lo}
$(Q_{cut}^{th})^2 = (Q_{cut}^{exp})^2 + Q_0^2$. 
In Fig. 2 we show predictions  for
the rapidity gap probability refering now to jets at resolution
$Q_{cut}^{exp}=\sqrt{y_{cut}} Q$ in the Durham algorithm;
we neglect the crossover effects which should be small at larger $Q_{cut}$. 
One can see the dramatic
rise of $f(\Delta y)$
at $\Delta y=4$ by about three orders of magnitudes if we replace hadrons
by jets at resolution 1 GeV. It will be interesting to verify this new
effect.   

\section{Conclusions, a Puzzle and a Physical Picture}
We have derived perturbative predictions for rapidity gap distributions 
using the two parameters $Q_0$ and $\Lambda$  from earlier
fits to the mean global particle multiplicity.
The  agreement with the SLD data is quite remarkable and in support of the
simple duality picture also in case of this new, 
partially exclusive, observable.   
It would be desirable to determine gap fractions from final states
with inclusion of neutral particles 
which would allow a more direct comparison with our calculations.
A crucial test of our picture is the strong dependence of the gap
distribution on the jet resolution.

Whereas the phenomenological description of the model is successful, 
the interpretation imposes a serious puzzle: for a large gap, take $\Delta
y = 3$, the first gluon in each hemisphere 
is emitted only after a mean  lifetime of about
10-20~f,  so the perturbative evolution is not disturbed, even if
the initial $q\overline q$ pair 
gets separated far beyond the typical confinement distance.\footnote{
We estimate the lifetime of the virtual quark radiating a gluon
with momenta $k,k_T$ as\cite{dkmt} 
$\tau\sim \frac{1}{m_q}\frac{k}{m_q}\sim \frac{k}{k_T^2}$ and take the
average of $\tau$ over $\omega'$ and $\Theta'$ as in (\ref{wp}) within the DLA.}

\begin{figure}[h,b,t]
\begin{center}  
\epsfxsize=7cm
\epsfbox{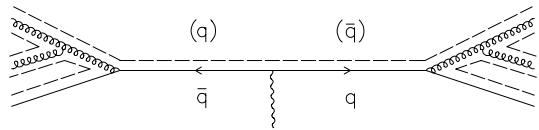} 
\end{center}                                                                
\caption{
Hadron final state emerging from $q\overline q$ with large
rapidity gap: the first gluon emission occurs far outside the confinement
region of size $\sim$1 f. A possible mechanism is colour blanching by 
soft $q\overline q$ pairs with $k_T\lesssim Q_0$ at all vertices.
\vspace{-0.3cm} 
 \label{fig:blanch}}
\end{figure}

As the mechanism for global colour blanching is not quantitatively known
we may consider phenomenological scenarios. 
The colour blanching could be mediated by ``gluers''\cite{dkmt} at $k_T\sim
Q_0$, they are expected to cause the production of hadrons with flat
rapidity plateau, 
already in absence of perturbative gluons. In our interpretation
there should not be
any associated real hadron production in the blanching process 
as the gap would be refilled. 
The confinement effects
must be weak enough so that the successful perturbative calculation 
of the gap rate is not  
 invalidated.  We consider the possibility that in the field of the
separating  $q\overline q$ pair (say, inside a tube with virtual partons of
$k_T<Q_0$) a 
very soft new $\overline q q$ pair is produced to ensure confinement
at usual distances, and further on at other vertices (Fig.  \ref{fig:blanch}). 
In a more specialized model the 
picture in Fig. \ref{fig:blanch} could be realized
 by effective hadronic vertices\cite{co}. At any rate,
it will be interesting to test further the predictions of the perturbative
analysis at low energy scales. 

\end{document}